\input harvmac \newcount\figno \figno=0 \def\fig#1#2#3{
\par\begingroup\parindent=0pt\leftskip=1cm\rightskip=1cm\parindent=0pt
\baselineskip=11pt \global\advance\figno by 1 \midinsert \epsfxsize=#3
\centerline{\epsfbox{#2}} \vskip 12pt {\bf Fig. \the\figno:} #1\par
\endinsert\endgroup\par } \def\figlabel#1{\xdef#1{\the\figno}}
\def\encadremath#1{\vbox{\hrule\hbox{\vrule\kern8pt\vbox{\kern8pt
\hbox{$\displaystyle #1$}\kern8pt} \kern8pt\vrule}\hrule}}

\overfullrule=0pt

%
\def\tilde{\widetilde}
\def\bar{\overline}

\font\zfont = cmss10 

\def\bigone{\hbox{1\kern -.23em {\rm l}}}
\def\ZZ{\hbox{\zfont Z\kern-.4emZ}}

\Title{hep-th/9807154, IASSNS-HEP-98/55}
{\vbox{\centerline{A Note on the Hitchin System  }
\medskip
\centerline{in a Background $B$-field}}}
\smallskip
\centerline{Hugo
Garc\'{\i}a-Compe\'an\foot{Also {\it Departamento de
F\'{\i}sica, Centro de Investigaci\'on y de Estudios Avanzados del IPN,
Apdo. Postal 14-740, 07000, M\'exico D.F., M\'exico} E-mail:
compean@sns.ias.edu}}
\smallskip
\centerline{\it School of Natural Sciences}
\centerline{\it Institute for Advanced Study}
\centerline{\it Olden Lane, Princeton, NJ 08540, USA}
\vskip .5truecm

\bigskip
\medskip
\vskip 1truecm
\noindent

The space of solutions to the Hitchin equations on the dual torus with punctures
determines the
Higgs branch of certain impurity theories. An alternative description of this
Higgs branch is provided, in terms of the proper deformation of Hitchin system with
deformation parameter
given by a $B$-field. For the dual torus minus the singular points
we construct
explicit solutions to the $B$-deformed Hitchin equations by reducing them to
principal chiral model equations and then using deformation quantization
methods.

\vskip 1truecm
\noindent

\noindent

\Date{07-1998}


\vfill
\break

\noindent
{\it Introduction}

\nref\witten{E. Witten, ``Bound States of Strings and $p$-Branes'',
Nucl. Phys. B {\bf 460} (1996) 299.}
\nref\alan{A. Connes, {\it Noncommutative Geometry}, Academic Press, 1994.}

Bound states of $N$ coincident parallel D-branes are described by a
SU$(N)$ supersymmetric gauge theory on the world-volume of the D-branes.
The scalar fields of this theory are matrices in the adjoint
representation of the gauge group SU$(N)$ and are interpreted as the
matrix transverse coordinates of the D-branes \witten. This gauge theory
possesses many features already found in gauge theories on noncommutative
spaces \alan.

Recently the study of matrix compactifications on tori has been
intensively worked out in the context of noncommutative geometry. The
consideration of a constant three form $C_{-ij}$ on the torus has shown to
deform the underlying gauge theory of toroidal compactification
\nref\connes{A. Connes, M.R. Douglas and A. Schwarz, ``Noncommutative
Geometry and Matrix Theory: Compactification on Tori'', hep-th/9711162.}
\nref\hull{ M. Douglas and C. Hull, ``D-Branes and Non-commutative
Torus'', hep-th/9711165.} \refs{\connes,\hull}. This deformation turns the
algebra of functions on the dual torus into the algebra with the Moyal
$*$-product defined in terms of the background $B$-field as a symplectic
form on the dual torus with its constant component as the deformation
parameter $\zeta_B$. By considering Type IIA theory with $N$ D0-branes on
${\bf T}^2$ in a background $B$-field one obtains the similar deformed
gauge theory \nref\cheung{ Y-K. E. Cheung and M. Krogh, ``Non-commutative
Geometry from 0-branes in a Background $B$-field, hep-th/9803031; T.
Kawano and K. Okuyama, ``Matrix Theory on Noncommutative Torus'',
hep-th/9803044.} \cheung.

\nref\eva{O. Aharony, M. Berkooz, S. Kachru, N. Seiberg and E. Silverstein,
``Matrix Description of Interacting Theories in Six Dimensions'',
hep-th/9707079.}
\nref\berkooz{O. Aharony, M. Berkooz and N. Seiberg,
``Light-Cone Description of (2,0) Superconformal Theories in Six Dimensions'',
hep-th/9712117.}
\nref\schwarz{N. Nekrasov and A. Schwarz,
``Instantons on Noncommutative ${\bf R}^4$, and (2,0) Superconformal
Six-dimensional Theory'', hep-th/9802068.}
\nref\micha{M. Berkooz, ``Non-local Field Theories and the Noncommutative
Torus'', hep-th/9802069.}

(2,0) field theories in six dimensions can be interpreted, in the light of
matrix theory, as a sigma model whose target space is the moduli space (with
some singularities corresponding to small instantons) of Yang-Mills (YM)
instantons on ${\bf R}^4$ \eva. A DLCQ description of these theories leads
to the introduction of the Fayet-Iliopoulos (FI) terms in the gauge theory in
order to resolve the singularities of the moduli space \berkooz. From the
space-time point of view FI parameters are interpreted as the constant
value of the background $B$-field. The resulting (2,0) theory can be
reinterpreted as a non-local field theory, with a finite non-locality
scale $\zeta$, describing YM instantons on noncommutative ${\bf R}^4$
\refs{\schwarz,\micha}. In \schwarz\ it was shown that the Higgs branch of
$N$ D0-branes inside $k$ D4-branes with D0 and D4 branes connected by the
expectation value of a $B$-field flux, parametrizes the instanton moduli
on the noncommutative ${\bf R}^4$ and thus a noncommutative version of the
ADHM construction of Yang-Mills instantons can be carried out. Generalizations
to the equivariant ADHM construction on noncommutative ALE
spaces and its comparison to Nakajima's description of instantons on ALE
spaces are given in \nref\laza{C.I. Lazaroiu, ``A Noncommutative-geometric
Interpretation of the Resolution of Equivariant Instanton Moduli Spaces'',
hep-th/9805132.} \laza.

The above system of D0 and D4-branes has also been recently discussed in
the context of impurity theories \nref\kapustin{A. Kapustin and S. Sethi,
``The Higgs Branch of Impurity Theories'', hep-th/9804027.} \kapustin.
There it was shown that the Higgs branch of impurity theories has a
hyper-K\"ahler structure and is given by the moduli space of
one and two-dimensional (compact) reductions of self-dual Yang-Mills equations.
The Nahm and Hitchin equations with impurity terms given by the fundamental
hypermultiplets are introduced by the longitudinal D4-branes. These results
have been used to solve elliptic models of ${\cal N}=2$ gauge theories
\nref\ed{ E. Witten, ``Solutions of Four-Dimensional Field Theories
Via M-Theory, Nucl. Phys. B {\bf 500} (1997) 3-42.} \ed\ via
compactification to three dimensions \nref\kapustint{A. Kapustin,
``Solutions of ${\cal N}=2$ Gauge Theories Via Compactification to Three
Dimensions'', hep-th/9804069.} \kapustint. The Coulomb branch of elliptic
models is mapped to the
Higgs branch of five-dimensional theories with three-dimensional
impurities. Thus its solution is given by solving a Hitchin system on
a Riemann surface with punctures
\kapustint.

Thus the Higgs branches of the mentioned theories do not receive quantum
corrections and they will be deformed under the
presence of the $B$-field \refs{\berkooz,\schwarz,\micha}.  In the present
note we argue that the Higgs branch of impurity theories, that is, the
moduli space of Hitchin equations with impurities will be deformed under
the presence of a background $B$-field. We will show that Moyal deformed
Hitchin equations can be solved explicitly through the BFFLS formalism
of deformation quantization \nref\bayen{ F. Bayen, M. Flato, C. Fronsdal,
A. Lichnerowicz and D. Sternheimer, ``Deformation Theory and Quantization
I ,II'', Ann. Phys. {\bf 111} (1978) 61, 111.} \bayen.

This paper is dedicated to Professor Jerzy F. Pleba\'nski on the occasion
of his 70th birthday. Although exact solutions of the Einstein equations have
been his constant preoccupation in physics he has also made some of the
leading contributions to self-dual gravity whose formalism was shown to be
relevant to describe the geometry of
${\cal N}=2$ strings.  Noncommutative deformation of the Hitchin equations
and their solutions are natural solutions of integrable systems present in
non-perturbative gauge theories coming from string theory and also from
matrix theory. I hope that the modest contribution to understanding these
systems presented here is appropriate on this occasion.

\vskip 1truecm

\noindent
{\it Impurity Vacua in a B-field}

In \kapustin\ it was shown that the Higgs branch of the bound state of $N$
D0-branes and $k$ D4-branes wrapped on a circle is given by the Nahm
equations with impurities determined by hypermultiplets coming from D0-D4 strings.
The corresponding system for a compactification of the D4-branes on ${\bf T}^2$ gives
the Hitchin equations with impurities localized at certain points on the dual torus $\widehat{\bf T}^2$
with real coordinates $\tilde{\sigma}$ and $\tilde{\sigma}'$

\eqn\hitchineq{ F_{z \bar{z}} + [\Phi, \bar{\Phi}] = {1 \over 2 R_1R_2}
\sum_{p=1}^{k} \big( Q^p \otimes Q^{*p} - \tilde{Q}^{*p}\otimes
\tilde{Q}^{p} \big)}

\eqn\diff{ \bar{D} \Phi = -  {1 \over 2 R_1R_2} \sum_{p=1}^{k}
\delta^2(\tilde{z} - \tilde{z}_p) Q^p \otimes \tilde{Q}^{p}}
where $F_{z \bar{z}} = \partial \bar{A} - \bar{\partial}A + [A,\bar{A}]$
and $\bar{D} = \bar{\partial} + [\bar{A}, \cdot]$ with $\partial =
\partial_z$, $\bar{\partial} = \partial_{\bar{z}}$, $A= A_z$ and $\bar{A}=
A_{\bar{z}}$.  Here $z=\tilde{\sigma} + i \tilde{\sigma}',$
$\bar{z}=\tilde{\sigma} - i \tilde{\sigma}',$ $\partial = {1\over
2}(\partial_{\tilde{\sigma}} - i \partial_{\tilde{\sigma}'})$ and $
\bar{\partial} = {1\over 2}(\partial_{\tilde{\sigma}} + i
\partial_{\tilde{\sigma}'}),$ $\Phi$ is a complex scalar field in the
adjoint representation of SU$(N)$ and $R_1,R_2$ are the radii of the
compact directions along the D4-branes. $A$ and $\Phi$ fields are functions
on $\widehat{\bf T}^2$ and they come from the adjoint hypermultiplets associated
to the D0-D0 strings. $Q$ and $\tilde{Q}$ are $k$ complex scalars belonging to
$k$ fundamental
hypermultiplets coming from D0-D4 strings, and $\tilde{z}_p =(z,\bar{z})_p$
are the positions of the $p$-th field $Q$ in the dual torus $\widehat{\bf T}^2.$
Scalar fields of $Q$ never become fields on the dual torus because they come from
longitudinal D4-branes.
Eqs. (1) and (2) should be understood  modulo gauge transformations and
the corresponding moduli space of solutions have no quantum corrections
and therefore the study of the Higgs branch lead to exact results using
only the classical equations (1) and (2).

Turning on a $B$-field flux on the impurity theory induces the presence
of FI terms $\xi_i$ such that the ${\cal D}$-flatness conditions are modified
to ${\cal D} = \xi_i$. Thus moduli space deformed by $B$-flux is again
interpreted as the description of YM instantons on noncommutative ${\bf
R}^2 \times \widehat{\bf T}^2$ \kapustint.

An alternative description of Higgs branch with a $B$-field can be done
through the Moyal deformation of self-dual YM equations \schwarz\ following the
lines of \refs{\connes,\cheung}. Thus
away from the singularities at the points $\tilde{z}_p$ and in the presence
of a $B$-field we have the Moyal deformed Hitchin equations on the
noncommutative dual torus $\widehat{\bf T}_B^2$ without punctures

\eqn\moyal{{F}_{z \bar{z}}(\tilde{z}) + \{\Phi(\tilde{z}),
 \bar{\Phi}(\tilde{z})\}_B = 0}

\eqn\moyalin{\bar{\partial} \Phi(\tilde{z}) + \{\bar{A}(\tilde{z}),
 \Phi(\tilde{z})\}_B = 0, \ \ \ \ {\partial} \bar{\Phi}(\tilde{z}) +
 \{{A}(\tilde{z}),
 \bar{\Phi}(\tilde{z})\}_B = 0, }
Here the Moyal bracket is defined by $\{f,g\}_B \equiv {1 \over i \zeta_B}
(f*g -g*f)$ for $f,g$ functions on the torus.
 The $*$-product is given by $f*g= f exp({i\over 2}\zeta_B \varepsilon_{ij}
\buildrel {\leftarrow}\over {\partial}_i \buildrel {\rightarrow}\over
{\partial}_j) g$ ($i,j=z,\bar{z})$ with $\zeta_B = B$, the deformation
parameter. The FI parameters $\xi$ are encoded in the $B$-deformation of
Eqs. (1) and (2). As one takes $B\to 0$ one
recovers straightforwardly system (1) and (2). This is equivalent to taking
 $\xi \to 0$.
 Eqs. (3) and (4) can be seen as
dimensional reduction of self-dual Yang-Mills equations on noncommutative
${\bf R}^2 \times \widehat{\bf T}^2$ to noncommutative $\widehat{\bf T}^2.$
From now on we will work with Eqs. (3) and (4) and in the next section we
attempt to find solutions for them.

\vfill
\break

\noindent
{\it Looking for Solutions of the Hitchin Equations in a B-field}

The Hitchin equations are two-dimensional reductions from self-dual Yang-Mills
theory \nref\hitchin{N.J. Hitchin, ``Self-duality Equations on a Riemann
Surface'', Proc. London Math. Soc. {\bf 55} (1987) 59.} \hitchin. These
equations are defined on any Riemann surface of genus $g$ and certain
marked points. The moduli space of Hitchin equations possesses an
hyper-K\"ahler structure. The original application of these equations was
to study the moduli space of stable vector bundles on an arbitrary Riemann
surface \hitchin.  Hitchin equations were also used brilliantly also to
gain more insight about this moduli space from the point of view of a Hamiltonian
integrable system
\nref\hh{N.J. Hitchin, ``Stable Bundles and Integrable Systems'', Duke
Math. J. {\bf 54} (1987) 91-114.} \refs{\hitchin,\hh}. Later Donaldson showed
that Hitchin
equations can be reduced to the study of twisted harmonic maps on a hyperbolic
space of negative curvature, and in this
context some solutions can be obtained \nref\donaldson{S.K. Donaldson,
``Twisted Harmonic Maps and the Self-duality Equations'', Proc. London
Math. Soc. {\bf 55} (1987) 127.} \donaldson.  Our goal in this section is
to obtain solutions for the Moyal $B$-deformed Hitchin equations (3) and (4). In
order to do it we first show the equivalence of Hitchin equations and the
principal chiral model at the classical level an then we use the BFFLS formalism
to look for explicit solutions \bayen.

\nref\dunne{G. Dunne, ``Chern-Simons Solitons,
Toda Theories and the Chiral Model'', Commun. Math. Phys. {\bf 150} (1992)
519.}

Following \refs{\hitchin,\donaldson} (see also \dunne) we define new connections ${\cal A} =
A - \Phi$ and $\bar{\cal A} = \bar{A} + \bar{\Phi}$. It can be easily shown
that ${\cal A}$ and $\bar{\cal A}$ are the components of a flat connection
on $\widehat{\bf T}_B^2$ if the Hitchin equations (3) and (4) are fulfilled.
That means ${\cal A}$ and $\bar{\cal A}$ satisfy

\eqn\flat{{\cal F}_{z \bar{z}}(\tilde{z}) = \partial \bar{\cal A}
(\tilde{z}) - \bar{\partial}{\cal A}(\tilde{z}) + \{{\cal A}(\tilde{z}),
\bar{\cal A}(\tilde{z}) \}_B = 0.}
Here we have used the existence of an harmonic map $g: \widehat{\bf
T}^2\to G_*,$ which satisfy ${\cal A}(\tilde{z}) = g^{- \buildrel
{*}\over{1}}(\tilde{z}) * \partial g(\tilde{z})$ and $\bar{\cal
A}(\tilde{z}) = g^{- \buildrel {*}\over{1}} (\tilde{z}) * \bar{\partial}
g(\tilde{z})$. $G_*$ is an infinite-dimensional Lie group which is defined
by $G_*=\{ g=g(\tilde{z}) \in C^{\infty}(\widehat{\bf T}^2); g* g^{-
\buildrel {*}\over{1}} = g^{- \buildrel {*}\over{1}}*g =1\}$ and $g^{-
\buildrel {*}\over{1}}$ is the inverse mapping in the group.  Defining
$H(\tilde{z})\equiv g(\tilde{z})* \Phi(\tilde{z}) * g^{- \buildrel
{*}\over{1}}(\tilde{z}),$ one can show that Hitchin's equations (3),(4) are
equivalent to the system

\eqn\ilis{g^{- \buildrel {*}\over{1}}(\tilde{z})
 *\bigg(\bar{\partial} H + {\partial}H -2 \{\bar{H},H\}_B\bigg)*
g(\tilde{z}) = 0,}

\eqn\prechiral{\bar{\partial} H(\tilde{z}) = \{\bar{H}(\tilde{z}),
H(\tilde{z}) \}_B, \ \ \ \ \ {\partial} \bar{H}(\tilde{z}) =
\{H(\tilde{z}),\bar{H}(\tilde{z}) \}_B.}

Furthermore one can define $J=2H$ and $\bar{J} = -2\bar{H}$ and the above
system is equivalent to the principal chiral model (PCM)

\eqn\flat{\partial \bar{J}(\tilde{z}) - \bar{\partial}J(\tilde{z}) +
\{J(\tilde{z}), \bar{J}(\tilde{z})\}_B = 0,}

\eqn\chiral{ \partial \bar{J}(\tilde{z}) + \bar{\partial} J(\tilde{z})
= 0}
with $J(\tilde{z}) = h^{- \buildrel{*}\over{1}}(\tilde{z}) * \partial
h(\tilde{z})$ and $\bar{J}(\tilde{z}) = h^{-
\buildrel{*}\over{1}}(\tilde{z}) * \bar{\partial} h(\tilde{z})$. These
equations can be derived from the Lagrangian

\eqn\lag{L_{PCM}= -{\zeta_B^2\over 2} \int d^2 \tilde{z} {\rm Tr}_N
\bigg(  h^{- \buildrel{*}\over{1}}(\tilde{z}) *
{\partial} h(\tilde{z}) *
 h^{- \buildrel{*}\over{1}}(\tilde{z}) *
\bar{\partial} h(\tilde{z}) \bigg).}

Equations (9) and (10) are precisely a suitable Moyal deformation of the PCM equations.
These equations are classically equivalent to the equations \nref\nappi{
C.R. Nappi, Phys. Rev. D {\bf 21} (1980) 418.} \nappi\

\eqn\gravity{i \partial \bar{\partial} \Theta(\tilde{z}) + {1\over 2}
\{ \partial{\Theta}(\tilde{z}), \bar{\partial}
\Theta(\tilde{z})\}_B=0}
where $J = - {1\over 2} \partial \Theta$ and $\bar{J}= {1\over 2}
\bar{\partial} \Theta$ with Lagrangian

\eqn\heaven{L_H = \int d^2 \tilde{z} {\rm Tr}_N \bigg( {1\over 2}(
\partial\Theta * \partial \Theta + \bar{\partial} \Theta *\bar{\partial}
\Theta)  + {2\over 3 }\Theta * \{\partial \Theta,\bar{\partial}\Theta\}_B
\bigg).}
From the quantum point of view the $\Theta$ model and the PCM are inequivalent because
both models are renormalized in a different way.
As we mentioned before the Higgs branch is determined exactly by classical
equations thus equivalence of Eqs. (8),(9) and (12),(13) we will use
is justified.

After having shown equivalence between Hitchin equations (3)(4) and the PCM
equations (9)(10) we now attempt to find solutions to the latter
equations. Before that it is convenient going back to real coordinates
$\tilde{\sigma}$.  Thus Eq. (12) is rewritten as

\eqn\realheaven{ \partial^2_{\tilde{\sigma}}\Theta +
\partial^2_{\tilde{\sigma}'}\Theta + \{\partial_{\tilde{\sigma}}\Theta,
\partial_{\tilde{\sigma}'}\Theta\}_B = 0} where $\Theta=
\Theta(\tilde{\sigma},\tilde{\sigma}')$.
This equation looks like the heavenly equation discussed in \nref\maciej{
J.F. Pleba\'nski, M. Przanowski and H.  Garc\'{\i}a-Compe\'an, Mod. Phys.
Lett.  A, {\bf 11} (1996) 663, hep-th/9509092.} \maciej. However in this
case $\Theta$ is a function on the dual torus and not on a
four-dimensional self-dual space.  Thus solutions of (14) do not determine
any self-dual metric. But one can still attempt to solve Eq. (14) using
the Weyl-Wigner-Moyal formalism. \nref\taylor{W. Taylor IV, ``D-brane
Field Theory on Compact Spaces'', Phys. Lett.  B {\bf 394} (1997) 283,
hep-th/9611042.} \nref\ganor{O.J.  Ganor, S. Ramgoolam and W. Taylor IV,
``Branes, Fluxes and Duality in Matrix Theory'', Nucl. Phys. B {\bf 492}
(1997) 191, hep-th/9611202.} The correspondence between matrices ${\bf
\Theta}$ and matrix-valued functions ${\Theta}(\tilde{\sigma})$ on the
dual torus is given by \refs{\taylor,\ganor,\connes}

$$\sigma_N^{-1}: Mat_N \to C^{\infty} (\widehat{\bf T}^2)$$

\eqn\weyl{\sigma_N^{-1}({\bf \Theta})= {\Theta}(\tilde{\sigma})  =
\sum_{(m,n) \in {\bf Z}^2} {\bf \Theta}_{m,n} exp\bigg( i({m\over
\widehat{R}_1} \tilde{\sigma} + {n\over \widehat{R}_2}
\tilde{\sigma}')\bigg)}
where $\widehat{R}_1 = {1 \over 2 \pi R_1}$, $\widehat{R}_2 = {1 \over 2
\pi R_2},$ $Mat_N$ is the set of $N \times N$ non-singular matrices
representing the Lie algebra su$(N)$ and $C^{\infty}(\widehat{\bf T}^2)$
is the set of smooth functions on the dual torus $\widehat{\bf T}^2$.

\nref\fairlie{D. Fairlie, P. Fletcher and C.K. Zachos, Phys. Lett. B {\bf 218}
(1989) 203; D. Fairlie and C.K. Zachos, Phys. Lett. B {\bf 224} (1989) 101;
T. Curtright, D. Fairlie and C. Zachos,  Phys. Lett. B {\bf 405} (1997) 37.}

On the other hand it is well known that the basis of the Lie algebra
su$(N)$ can be seen as a two-indices infinite algebra. The elements of
this basis are denoted by $exp\big( i({m\over \widehat{R}_1}
\tilde{\sigma} + {n\over \widehat{R}_2}\tilde{\sigma}')\big)$ and they
satisfy the two-indices infinite Lie algebra \fairlie\

\eqn\algebra{ [L_{\bf m}, L_{\bf n}] = {N \over \pi} {\rm sin} \big( {\pi
\over N} {\bf m} \times {\bf n} \big) L_{{\bf m} + {\bf n} \ \ {\rm mod} \
\ N{\bf q}},}
where ${\bf m}=(m_1,m_2)$, ${\bf n}=(n_1,n_2)$ and ${\bf m} \times {\bf n}
:= m_1n_2 - m_2 n_1$. The large $N$ limit ($N \to \infty$) of algebra
\algebra\ gives the area-preserving diffeomorphism algebra
sdiff$(\widehat{\bf T}^2).$

The correspondence \weyl\ can be seen as the composition of two mappings.
The first one is a Lie algebra representation of su$(N)$ (for {\it finite}
$N$)  into a Lie algebra $\widehat{\cal G}$ of self-adjoint operators
acting on the Hilbert space $L^2({\bf R}),$ given by

\eqn\une{{\bf \Psi}: {\rm su}(N) \to \hat{\cal G},
 \ \ \ \ \ {\bf \Theta} \mapsto {\bf \Psi}({\bf \Theta}) := \widehat{\bf
\Theta}.}
The second mapping is a genuine Weyl correspondence ${\cal W}^{-1}$ which
establishes a one to one correspondence between the algebra ${\cal B}$ of
self-adjoint linear operators acting on $L^2({\bf R})$ and the space of
real smooth functions $C^{\infty}(\widehat{\bf T}^2)$ where $\widehat{\bf
T}^2$ is seen as the classical phase-space. This correspondence $ {\cal
W}^{-1}: {\cal B} \to C^{\infty}(\widehat{\bf T}^2)$ is given by

\eqn\corresp{\Theta(\tilde{\sigma},\tilde{\sigma}';\zeta_B) \equiv
{\cal W}^{-1}(\widehat{\bf \Theta}) :=
\int_{- \infty}^{\infty} <\tilde{\sigma} - {\xi \over 2}|
\widehat{\bf \Theta}|\tilde{\sigma} + {\xi \over
2}> {\rm exp}\big( {i \over \zeta_B} \xi \tilde{\sigma}' \big) d\xi,}
for all $\widehat{\bf \Theta} \in {\cal B}$ and ${\Theta} \in C^{\infty}
(\widehat{\bf T}^2).$ Thus from the identification of ${\cal B}$ with
$\hat{\cal G}$, it follows that the correspondence $\sigma_N^{-1}$ is
equal to the map composition $\sigma^{-1}_N = {\cal W}^{-1} \circ {\bf
\Psi}$ for finite $N$ and it is actually a Lie algebra isomorphism.

In Ref. \maciej\ it was shown how to solve Eq.(14) for the case of
self-dual gravity. In what follows we apply the same method to find
solutions for this equation. We set $N=2$, obtaining $2\times 2$
matrices forming the Lie algebra su$(2)$. Thus matrices ${\bf \Theta}$ can be
expanded as

\eqn\suma{ {\bf \Theta} = \sum_{a=1}^{3}\theta_a \tau_a} where $\tau_a$
$(a=1,2,3)$ constitutes a basis of su$(2)$ and $\theta_a$ are some constant
numbers. Eq. (17) leads to

\eqn\iso{\widehat{\bf \Theta} = \sum_{a=1}^3 \theta_a {\bf \Psi}(\tau_a)}
where ${\bf \Psi}(\tau_a)$ $(a=1,2,3)$ are a basis for the Lie subalgebra
$\widehat{\rm su}(2)$ of unitary operators and are given by:
${\bf \Psi}(\tau_1)= i \alpha \widehat{\tilde{\sigma}}' + {1 \over 2 \zeta_B}
(\widehat{\tilde{\sigma}}'^2 -1)\widehat{\tilde{\sigma}}$,
${\bf \Psi}(\tau_2)= - \alpha \widehat{\tilde{\sigma}}' + {i \over 2
\zeta_B}
(\widehat{\tilde{\sigma}}'^2 + 1)\widehat{\tilde{\sigma}}$ and
${\bf \Psi}(\tau_3) = -i \alpha \widehat{\bf 1} - {1 \over \zeta_B}
\widehat{\tilde{\sigma}}'\widehat{\tilde{\sigma}}.$ Inserting equations for
${\bf \Psi}(\tau_a)$ into Eq. (20) and then using (18) we obtain finally the
solution

\eqn\solution{ \Theta(\tilde{\sigma},\tilde{\sigma}') = {i\over 2}
\theta_1 \tilde{\sigma} (\tilde{\sigma}'^2 -1)  - {1\over 2}
\theta_2 \tilde{\sigma} (\tilde{\sigma}'^2 +1) - i \theta_3 \tilde{\sigma}
\tilde{\sigma}' + B \cdot (\alpha+{1\over 2})(-\theta_1 \tilde{\sigma}' -
i \theta_2 \tilde{\sigma}' + \theta_3).}
When $\zeta_B( =B )\to 0$ we get the simple solution

\eqn\cero{ \Theta_0(\tilde{\sigma},\tilde{\sigma}') = {i\over 2}
\theta_1 \tilde{\sigma} (\tilde{\sigma}'^2 -1)  - {1\over 2}
\theta_2\tilde{\sigma}(\tilde{\sigma}'^2 +1) - i \theta_3 \tilde{\sigma}
\tilde{\sigma}'}
where $\Theta_0$ is the lower term of the series $\Theta = \Theta_0 +
\sum_{n=1}^{\infty} \zeta^n_B \Theta_n$.

Thus Hitchin's equations can be solved by reducing them to the PCM
equations and then using the WWM formalism. Solutions for the Hitchin equations
in a $B$-field depend explicitly on the deformation parameter $B$ as we
have shown in Eq. \solution. In this sense the moduli space of deformed
Hitchin equations is deformed and therefore the Higgs branch will also be
deformed. The same procedure can be carried over to the Higgs branch given
by the solutions of Nahm's equations away from impurities \kapustin. In
that case the presence of the background $B$-field turns the Nahm equations
into

\nref\hugo{H. Garc\'{\i}a-Compe\'an and J.F. Pleba\'nski, ``On the
Weyl-Wigner-Moyal Description of Nahm Equations'', Phys. Lett. A {\bf 234}
(1997) 5.}

\eqn\nahm{ {dT_i\over ds} + \{T_0,T_i\}_B + {1 \over 2}\varepsilon_{ijk}
\{T_j,T_k\}_B=0.} Solutions of these equations can be obtained
straighforwardly following the lines of \hugo.

\nref\giddings{S.B. Giddings, F. Hacquebord and H. Verlinde,
``High Energy Scattering and D-Pair Creation in
Matrix String Theory'', hep-th/9804121.}

\nref\bonora{G. Bonelli, L. Bonora and F. Nesti, ``Matrix String Theory,
2D SYM Instantons and Affine Toda Systems'', hep-th/9805071.}

Finally Hitchin equations on the cylinder ${\bf R} \times{\bf S}^1$ can be
obtained from matrix string theory compactified on ${\bf S}^1$
\refs{\giddings,\bonora}. Hitchin equations results form the BPS condition
in the supersymmetry transformations.  Compactification on a further
circle ${\bf S}^1$ one can find the equivalence with a supersymmetric
Yang-Mills theory on a $2+1$-dimensional spacetime ${\bf R} \times
\widehat{\bf T}^2$. If one can include a background $B$-field in this
picture, and check the BPS condition one obtains $B$-deformed Hitchin equations
on a dual torus $\widehat{\bf T}^2_B$ of the type (3)(4). Solutions of these
systems following the lines of the present paper would be an alternative than the
Toda equation method discussed in \bonora.

\vskip 3truecm
\centerline{\bf Acknowledgments}

I would like to thank A.  G\"uijosa, A. Kapustin, A.M. Uranga and Y-S. Wu
for useful discussions and comments.  This work is supported by a
Postdoctoral CONACyT fellowship under the program {\it Programa de
Posdoctorantes: Estancias Posdoctorales en el Extranjero para Graduados en
Instituciones Nacionales 1997-1998}. It is a pleasure to thank E. Witten for
hospitality in the Institute for Advanced Study.

\listrefs

\end